# The characterization of Co-nanoparticles supported on graphene

P. Bazylewski[1], D. W. Boukhvalov[2,6], A. I. Kukharenko[3,4], E. Z. Kurmaev[3,4], A. Hunt[1], A. Moewes[1], Y. H. Lee[5], S. O. Cholakh[4] and G. S. Chang[1]

[1]*Department of Physics and Engineering Physics, University of Saskatchewan, 116 Science Place, Saskatoon, SK, S7N 5E2, Canada*

[2]*Department of Chemistry, Hanyang University, 17 Haengdang-dong, Seongdong-gu, Seoul 133-791, Korea*

[3]*M.N. Mikheev Institute of Metal Physics, Russian Academy of Sciences-Ural Division, 620990 Yekaterinburg, Russia*

[4]*Ural Federal University, 19 Mira Str., 620002 Yekaterinburg, Russia*

[5] *Graphene Center, Samsung Advanced Institute of Technology, Sungkyunkwan University, Suwon 440-746, Republic of Korea*

[6]*Theoretical Physics and Applied Mathematics Department, Ural Federal University, Mira Street 19, 620002 Ekaterinburg, Russia*

## ABSTRACT

The results of density functional theory calculations and measurements using X-ray photoelectron spectroscopy of Co-nanoparticles dispersed on graphene/Cu are presented. It is found that for low cobalt thickness (0.02 nm - 0.06 nm) the Co forms islands distributed non-homogeneously which are strongly oxidized under exposure to air to form cobalt oxides. At greater thicknesses up to 2 nm the upper Co-layers are similarly oxidized whereas the lower layers contacting the graphene remain metallic. The measurements indicate a $Co^{2+}$ oxidation state with no evidence of a 3+ state appearing at any Co thickness, consistent with CoO and $Co[OH]_2$. The results show that thicker Co (2nm) coverage induces the formation of a protective oxide layer while providing the magnetic properties of Co nanoparticles.

**INTRODUCTION**

Transition-metal nanoparticles assembled on graphene have garnered great interest in recent years due to their potentially useful properties in many applications. Transition and noble-metal nanostructures on graphene have been shown to be efficient as supported catalysts [1,2]. In particular nanoparticles such as cobalt on graphene have been investigated for electrochemical oxygen reduction reactions (ORR) to replace more expensive catalysts such as Pt [1]. Platinum-based materials are known to be the most active catalysts for ORR in both acidic and alkaline conditions, but this fact is offset by high cost and limited stability [3]. Cobalt-based materials such as spinel oxides have been investigated as a potential alternative for electrocatalysts due to their high activity and stability, cost effectiveness, and relatively simple preparation [4]. Studies have shown that for such applications the addition of graphene or graphene oxide as a support can enhance catalytic activity by encouraging optimal electrical and chemical coupling between the nanoparticles, depending on the size and distribution of the particles [2].

Other technological interests concern the growth of uniform magnetic metal films for use as spin filters in spintronic devices, or as thermally stable magnetic islands placed in high density for data storage applications [5,6]. The study of metals on graphene is also important for a better understanding of the quality of metal contacts on graphene, which is very critical for optimal performance of graphene-based electronic devices [7]. Sufficiently low contact resistivity ($10^{-9}$ $\Omega \cdot cm^2$) for use in miniature graphene-based devices such as field-effect transistors has been shown to be difficult to achieve between metals and graphene because conduction occurs mainly through the edges of the metal contact, resulting in generally higher contact resistance with bulk electrodes [8]. For this reason, contacts constructed from successive layers of metal nanoparticles are more desirable to improve electrical contact. Magnetoelectric effects in Co/graphene systems have also been predicted as a result of exchange coupling between cobalt clusters placed on a graphene sheet [9]. First principles calculations have shown that a large exchange coupling, which can be ferromagnetic (FM) or antiferromagnetic (AFM) depending on the graphene site, is possible between Co islands on graphene or between a Co top layer and Co substrate separated by graphene [10]. The exchange coupling strength between metal clusters as well as the graphene charge carrier concentration are tunable using proper gating and may open the possibility of controllable magnetoelectric effects in Co/graphene systems [11]. The realization of these systems however requires precise control of cluster size and distribution; to achieve the desired FM of AFM coupling the clusters must be appropriately distributed on opposing graphene sublattices. The full characterization of such materials on an atomic and electronic level is highly desirable to understand the electronic properties of such metal/graphene composites and will contribute to further development of the applications mentioned above.

In the present paper, we have employed X-ray spectroscopic techniques and density functional theory (DFT) calculations to investigate the local atomic and electronic structure of Co nanoparticles deposited on a graphene/Cu substrate. To investigate the Co oxidation state and local bonding environment, X-ray photoelectron spectroscopy (XPS) measurements (O and C 1$s$, Co and Cu 2$p$ core levels and valence bands) were performed for Co/graphene/Cu systems with full and partial Co coverage on the graphene surface. The obtained results are compared with DFT calculations of the formation energies for oxygen adsorption depending on the number of upper Co layers and the electronic structure of a CoO/Co/graphene/Cu system. The results show for a relatively thick Co layer (thickness of 2 nm), the upper Co layers (70% of full thickness) are oxidized whereas the lower layers (30%) remain in a metallic state. At lower thicknesses, the Co layer is almost entirely oxidized into $Co^{2+}$ oxide species of CoO and Co[OH]$_2$. Below 0.06 nm only CoO is formed, with Co[OH]$_2$ becoming dominant as thickness increases up to 2 nm. These results demonstrate that AFM CoO can be formed on the graphene surface using carefully controlled deposition, producing a CoO/graphene system ideal for exchange coupling interactions, and use in spintronic devices.

## 2. EXPERIMENTAL

### 2.1. Sample Preparation

Graphene (Gr) was grown on mechanically and chemically polished Cu foil (99.96%, 100 μm-thick foil purchased from Nilaco) using atmospheric pressure chemical vapor deposition (APCVD) system (see Ref. [12] for details). The composite samples fabricated on graphene consisted of a material stack of the form Co/Gr/(Cu foil) with the Co layers deposited by physical vapour deposition (PVD) using Co powder (Sigma Aldrich, 99.995%) as the source material in a tungsten boat. The thicknesses reported are as measured *in situ* by a quartz crystal monitor, with the deposition rate maintained at 0.02 Å/s. The prepared samples were immediately stored and transported under light vacuum ($10^{-3}$ torr), and cleaved into smaller pieces prior to XPS measurements.

### 2.2. XPS Measurements

The XPS measurements were performed using a PHI XPS Versaprobe 5000 spectrometer (ULVAC-Physical Electronics, USA). The energy resolution was $\Delta E \leq 0.5$ eV for the Al *K*α excitation (1486.6 eV) and the pressure in the analysis chamber was maintained below $10^{-9}$ torr. The dual-channel neutralizer (ULVAC-PHI patent) was applied in order to compensate the local charging of the sample due to the loss of photoelectrons during XPS measurements. After the Co/Gr/Cu samples were kept in the preparation chamber for 24 hours prior to measurements, they were introduced into the analysis chamber and examined the presence of any possible micro-impurities using chemical state mapping mode. If the micro-impurities were detected, the sample was replaced from the reserved batch. The core-level and valence-band XPS spectra were recorded with Al $K\alpha$ radiation using a 100 μm spot and X-ray power load on the sample of less than 25 W. Typical signal to noise ratios were more than 10000/3.

**2.3 SIESTA Calculation Details**

The pseudo-potential code SIESTA [13] was used for DFT calculations, following a similar procedure found in previous papers [14, 15]. All calculations were based on the local density approximation (LDA) [16], which is feasible for the modelling of Co on graphene over a Cu substrate (see discussion in Ref. [14]). Full optimization of the atomic positions was performed where the wave functions were expanded with a double-$\zeta$ plus polarization basis of localized orbitals for all species. The force and total energy of the systems were optimized with an accuracy of 0.04 eV/Å and 1 meV, respectively. All calculations were carried out with an energy mesh cut-off of 360 Ry and a **k**-point mesh of 4×4×2 in the Monkhorst-Pack scheme [17]. The chemisorption energies for oxidation were calculated using a standard equation: $E_{chem} = E_{host+O} - (E_{host} + E_{O2})$ where $E_{host}$ is the total energy of the system before adsorption of oxygen atoms and $E_{O2}$ is the total energy of molecular oxygen in a triplet state. Due to the size of the formation energies under study (several eV), small perturbations such as zero-point energy (on the order of less than 0.2 eV) were not taken into account. It should be noted that this study is focused on the oxidation mechanism of Co as it depends on the thickness. For this reason the calculations were subject to the following conditions: (i) the size of Co clusters was limited by limiting the size of supercell, (ii) the thickness of cobalt layers was limited to a few layers required for 3D cluster formation, (iii) defects and grain boundaries in the graphene and cobalt were not considered, (iv) limited oxidation of the graphene intrinsically present was not included, (v) from results presented in Refs. 14 and 15, the lattice mismatch between graphene and the copper substrate does not significantly affect the chemical properties of this system, and thus was not considered in the calculations.

The oxidation of Co/Gr/Cu was simulated using the following model structures: (*a*) planar Co clusters of 1, 3, or 7 atoms on the Gr/Cu substrate consisting of 6 *fcc* Cu layers (16 Cu atoms in each layer) covered with a layer of graphene containing 32 C atoms (6, 18 or 43% of the Co coverage) and (*b*) 1, 2, 4, or 6 layers of Co on the same Gr/Cu structure. Some examples of model structures are illustrated in Fig. 1. A simulation of the oxidation is produced by placing an $O_2$ molecule close to the Co atoms, which is then decomposed at the surface to form bonds with Co atoms.

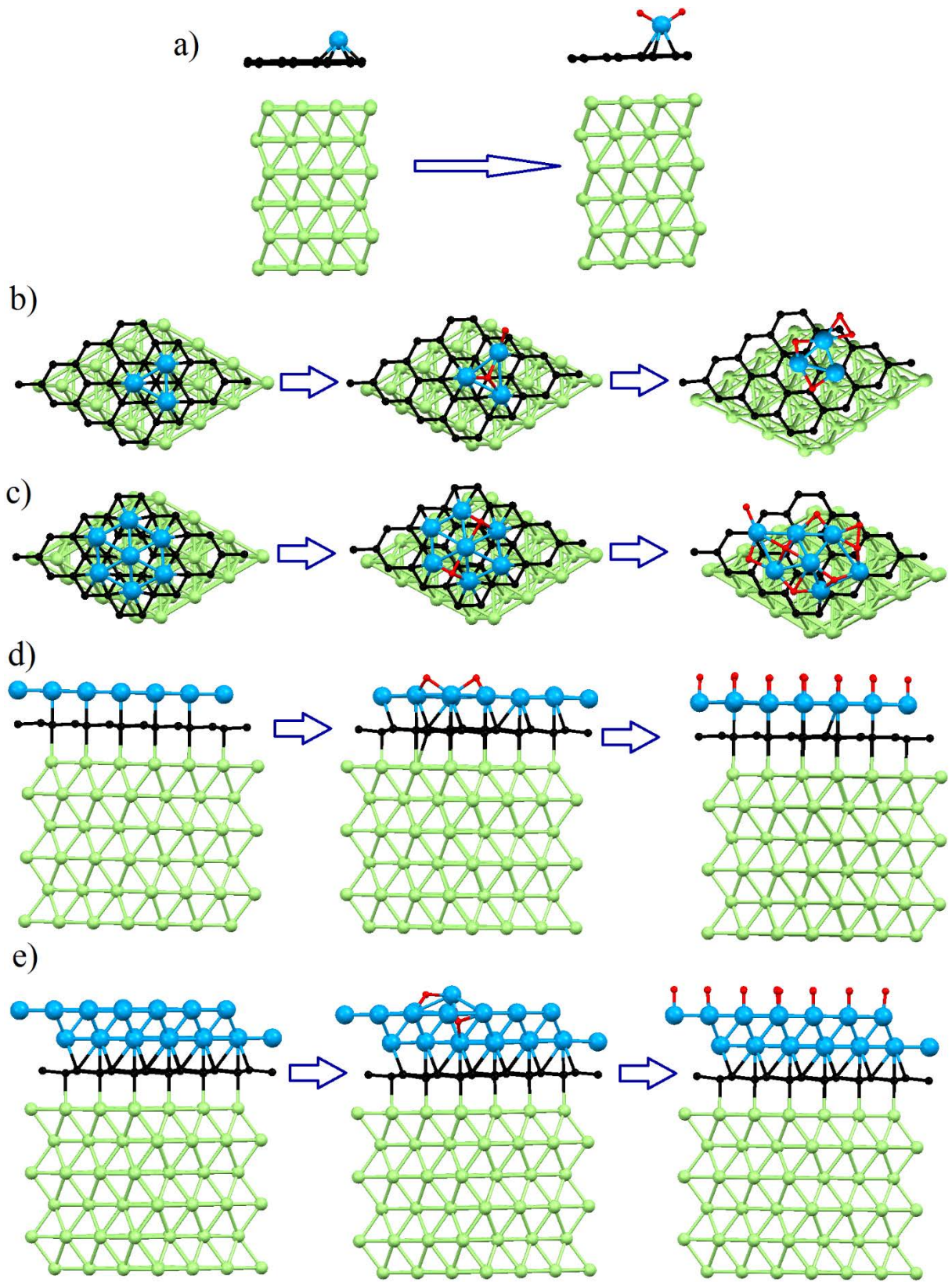

**FIG. 1.** Optimized atomic structures for initial, intermediate and final steps of the oxidation of a Co monoatom. (a) planar clusters, (b) monolayer, (b) bilayer, and (d and e) for graphene on a copper substrate.

## 3. RESULTS AND DISCUSSION

### 3.1 XPS Measurements and Analysis

The XPS survey spectra of Co-deposited Gr/Cu composites normalized to the Cu 2*p* XPS line of Cu foil are presented in Fig. 2. The XPS spectra show no additional impurities presented in all Co/Gr/Cu samples and the intensity of Co 2*p* line increases with increasing Co content. The oxygen content is also observed to increase in proportion to the Co thickness deposited, indicating formation of Co, C, or Cu oxides. To examine the oxygen content in graphene, further analysis was performed at the C and O 1*s* lines using a peak fitting.

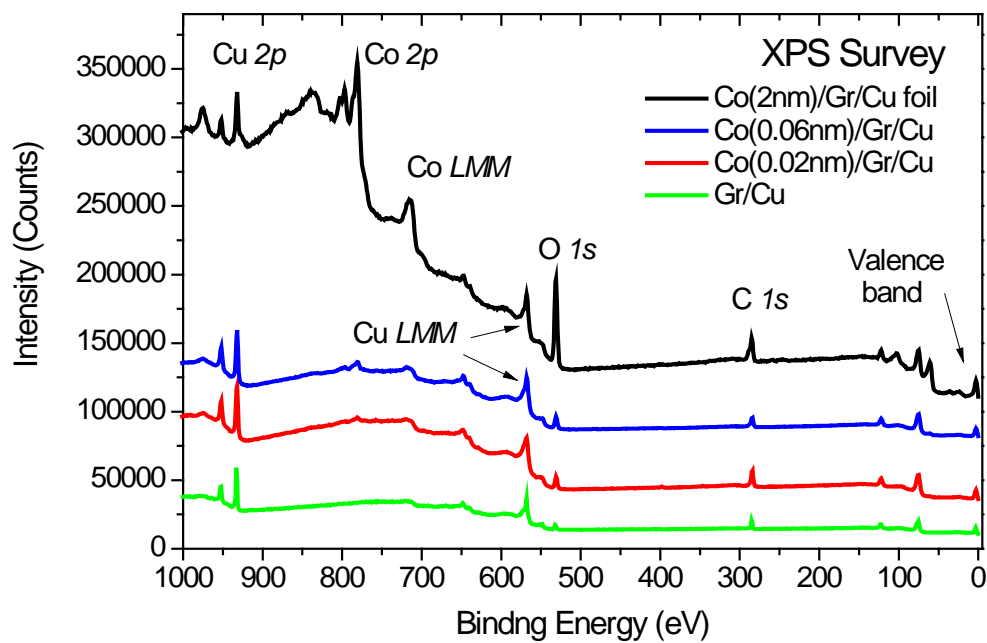

**FIG 2**. XPS survey spectra of Co(0.02 nm)/Gr/Cu, Co(0.06 nm)/Gr/Cu and Co(2 nm)/Gr/Cu (spectra are normalized to the Cu 2*p* intensity).

Peak fitting using Voigt functions was accomplished to identify potential oxygen functional groups present by considering the possible functional groups and their energy locations as reported in the literature for graphene oxide systems [18,19]. As seen in Fig. 3, C=O (288.6 eV) and C-O (285.8 eV) bonds are formed and increase with Co thickness, as well as carbonate ($CO_3^{-2}$) at higher thickness. These groups are typical of graphene oxide which can form epoxide (C-O-C), hydroxyl (C-OH), carbonyl (C=O) and carboxyl (COOH) both in the graphene basal plane and at defects or edges depending on the group. Some oxide groups (C=O and C-OH) are present in low concentration on as-prepared Gr/Cu, which can be attributed to native oxidation or adsorbed water. From the peak intensities at the O 1$s$ core level, the OH groups forming at higher Co thickness are not only due to C-OH bonding, but also formation of Co[OH]$_2$. There may also be a contribution from cobalt oxides at the O 1$s$ peaks in the range of 529-533 eV, but they could not be resolved using peak fitting [20].

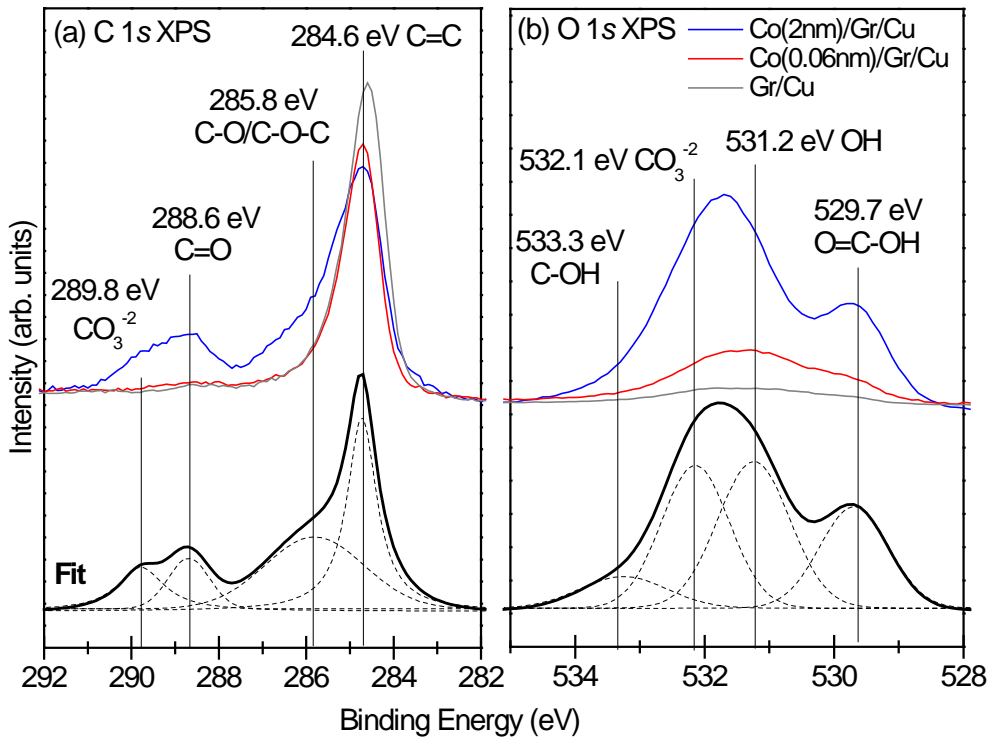

**FIG. 3**. (a) C1$s$ and (b) O 1$s$ core-level XPS spectra of Co/ Gr/Cu samples. Peak fitting was performed using Voigt functions on Co(2 nm)/Gr/Cu spectrum.

From these edges, the oxygen content increases proportionally to the thickness of deposited Co. This indicates that primarily Co oxides are forming in addition to oxidation of graphene. To further investigate the oxygen functionalization, the valence band (VB) spectra of these samples were measured and the results are presented in Fig.

4. By comparison to Co and Cu metal references as well as CoO powder, a contribution to the VB edge is apparent when Co is added. Additional electronic states corresponding to Co appear in the VB for all Co thicknesses. Metallic Co is also present, but only appears above approximately 1 nm of Co thickness (not shown).

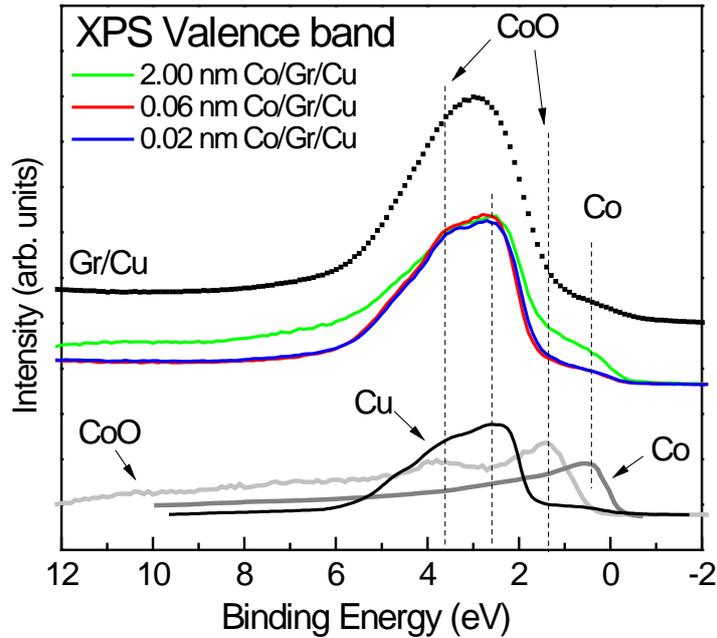

**FIG. 4**. Comparison of valence band XPS spectra of Co(0.02 nm)/Gr/Cu, Co(0.06 nm)/Gr/Cu and Co(2 nm)/Gr/Cu samples with reference spectra of Cu metal [13], Co metal [14] and CoO [14]. Some spectra have been vertically translated for clarity.

From comparison to Gr/Cu, a splitting of the main VB peak is observed after Co is added. The spectral feature at about 2.6 eV is contributed mainly from Cu $3d$ states, as is typical for Cu metal [21], because of the high photoionization cross-section of Cu $3d$ states for Al $K\alpha$ radiation compared to those of C and O $2p$ states [14]. Despite these spectral contribution from Cu and CoO, we cannot exclude the presence of a Co[OH]$_2$ phase because the VB spectrum of this material is very similar to that of CoO with the same divalent ionic state of Co (Co$^{2+}$). Therefore the Co $2p$ XPS is examined further, as detailed in Fig. 5.

Comparison of Co/Gr/Cu samples with Co islands (0.02 and 0.06 nm of Co thickness) to Co[OH]$_2$ reference shows good agreement with the lineshape and charge transfer satellites (S$_1$, S$_2$) which are typical for spectra of divalent Co [12]. The 2$p_{3/2}$ feature located at 780.7 eV and the associated 2$p_{1/2}$ peak at 796.8 eV exhibit an energy splitting (ΔE = 15.9 eV) which is also consistent with a Co$^{2+}$ state [22,23]. The spectrum of Co(2 nm)/Gr/Cu is somewhat more complicated, which is attributed to a superposition of Co$^0$ and Co$^{2+}$ signals in the ratio 3:7 (see Fig. 6).

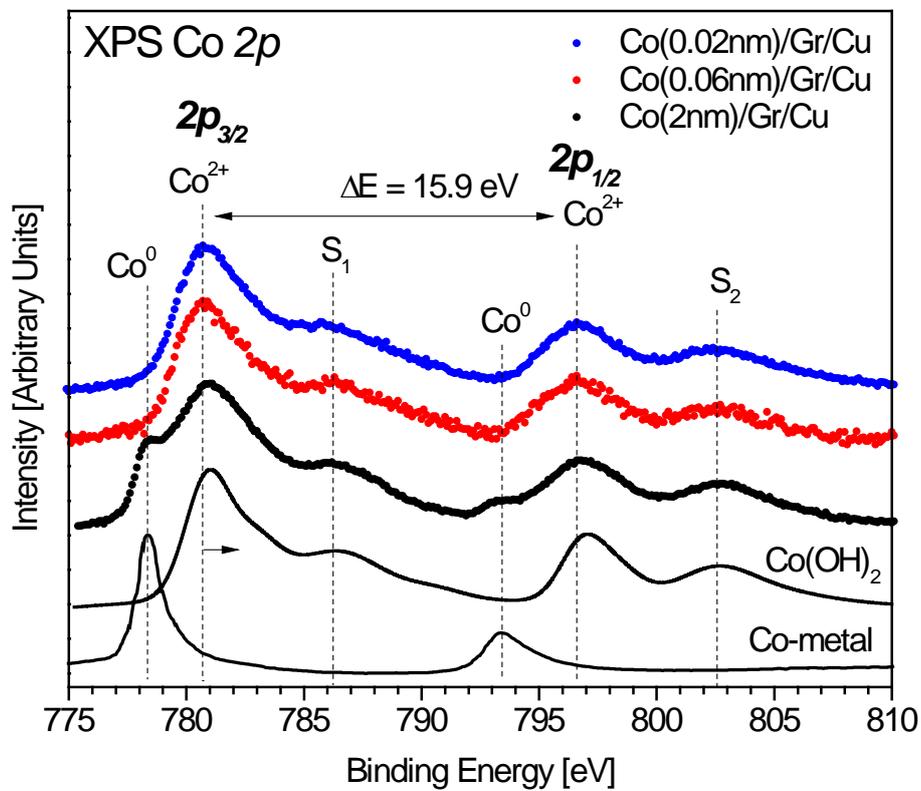

**FIG. 5**. Comparison of Co 2$p$ core-level XPS spectra of Co-coated Gr/Cu systems with reference spectra of Co metal and Co(OH)$_2$ [6].

It should be noted that there is also a shift of Co$^{2+}$ features between thin and thick Co layers. The Co 2$p_{3/2}$ peak present at 780.7 eV for 0.02-0.06 nm of Co, is observed to shift to higher energy (781.1 eV) for 2 nm of Co, causing the 2$p_{3/2,1/2}$ splitting to be reduced to 15.8 eV. This observed shift of the Co$^{2+}$ feature to higher energy is consistent with initial formation of mostly CoO, converting to Co[OH]$_2$ which has a slightly higher binding energy (about 0.5 -1 eV) [24].

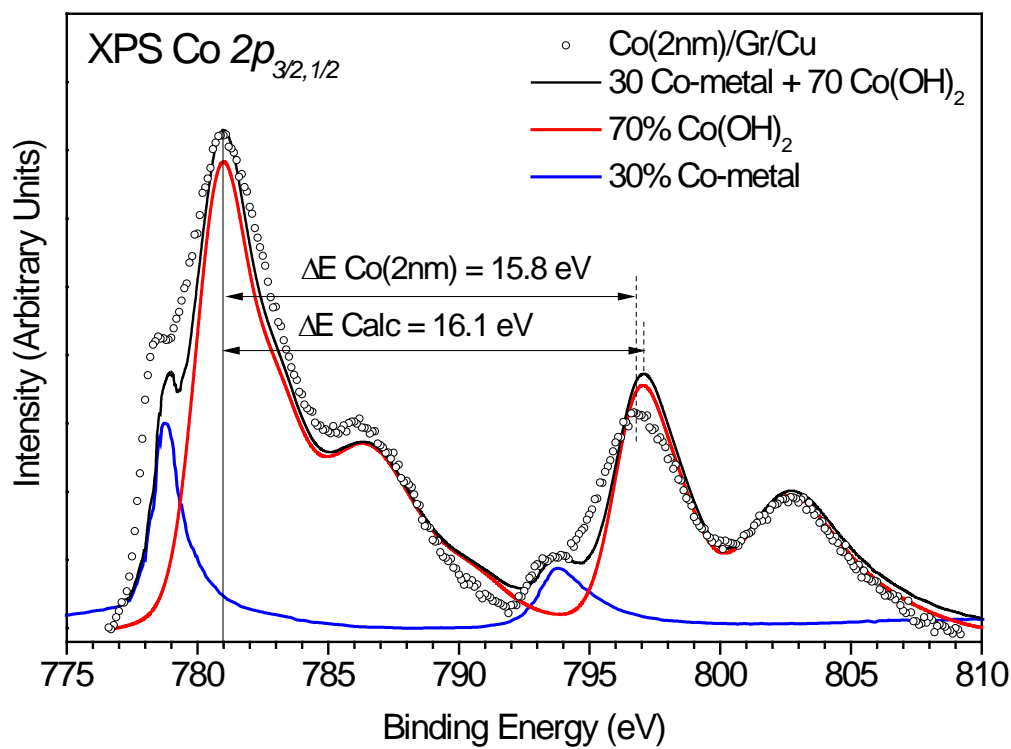

**FIG. 6.** Comparison of Co 2$p_{3/2,1/2}$ core-level XPS spectrum of Co(2nm)/Gr/Cu with a superposed spectrum with Co(OH)$_2$ and Co metal reference spectra.

The presence of a satellite feature (Co$^0$) at about 778.3 eV for a thick Co layer on graphene suggests the presence of Co metal, so the composition of Co(2 nm)/Gr/Cu is estimated by superposing the reference spectra of Co metal and Co[OH]$_2$. As shown in Fig. 6, the calculated spectrum of 30% Co metal and 70% Co[OH]$_2$ has very good agreement with the lineshape, but slightly larger peak splitting of 16.1 eV compared to the experimental value of 15.8 eV. The overall results confirm the structure of the thin and thick Co layers on graphene. At low Co thickness, the Co islands form primarily CoO with no evidence of metallic Co. With increasing Co thickness, the CoO phase is partially converted to Co[OH]$_2$, and for 2 nm-thick Co the bottom layers in contact with the graphene remain metallic Co.

### 3.2 SIESTA Calculation Results

For verification of the mode of oxidation discussed above, the Co/Gr/Cu system was also theoretically investigated using DFT calculations concerning the interaction of oxygen with various Co structures on a Gr/Cu substrate in an effort to understand the experimental results. In Fig. 7a, one can see that the spectral features contributed from Co, CoO and Cu in total density of states (TDOS) are fairly consistent with the experimental VB spectrum of Co(2 nm)/Gr/Cu sample presented in Fig. 4. The contribution to the partial density of states (PDOS) from the Co metal dominates near the Fermi level (Fig. 7b). Similarly Co and O atoms in the upper CoO layer also contribute to states near the Fermi level with other deep-valence states around −4~5 eV (Fig. 7.c). Contributions from metallic Co and CoO near -5 eV are in quantitative agreement with experimental results (Fig. 4). After Co deposition, the calculations indicate the VB character is largely determined by the amount of Co and cobalt oxides that are formed. The spin magnetic moments of Co ions in metallic layers are found to be ferromagnetically coupled within metallic layers with the values of exchange interactions on the order of 600 K in the *xy*-plane and 300 K between cobalt layers. The oxidized cobalt layer is also coupled with the unoxidized cobalt substrate ferromagnetically, with the value of this exchange interaction about 270 K. The magnetic moment values are significantly different between CoO and Co layers: 0.38 $\mu_B$/atom for the CoO layer and 2.11~1.92 $\mu_B$/atom for the metallic layers of Co. Taking into account on-site Coulomb repulsion contributes approximately a 10% change in the values of magnetic moments and exchange interactions in thr CoO-layer as compared to pristine CoO. [25].

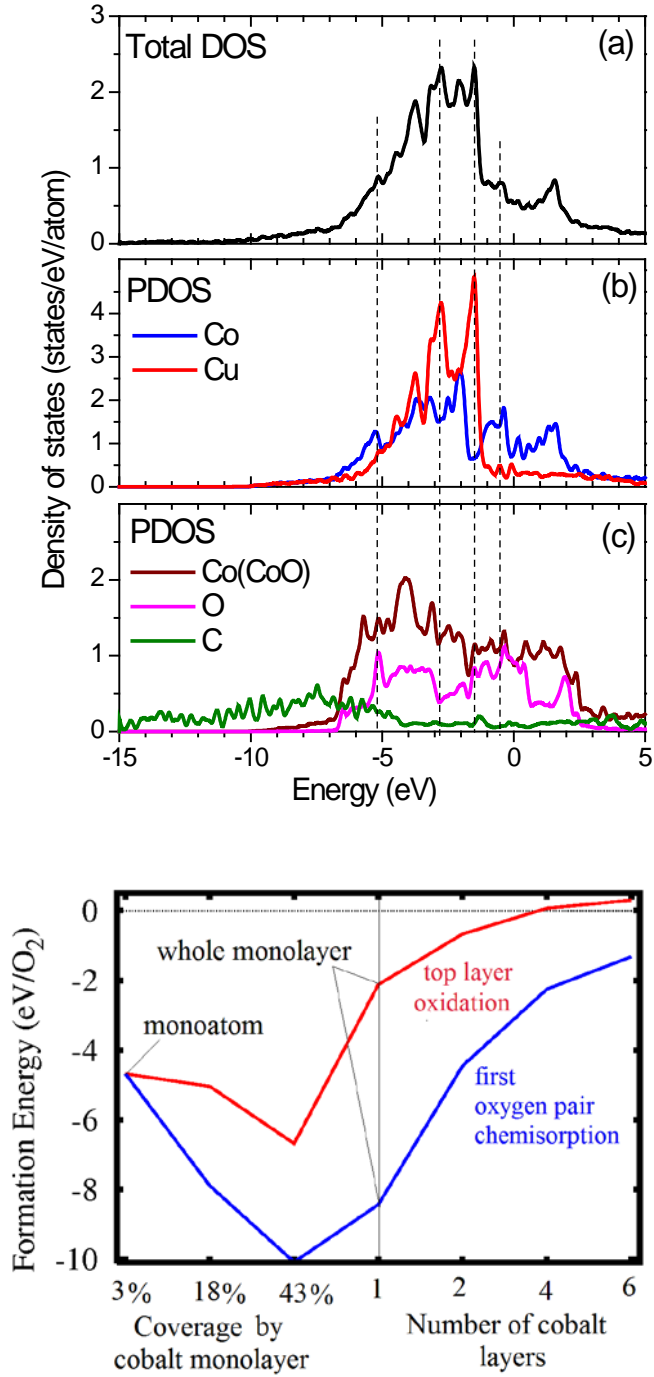

**FIG. 7.** (Top) (a) Calculated TDOS and (b and c) PDOS for a CoO/Co/Gr/Cu system. (Bottom) Calculated formation energies of adsorbed oxygen atoms with respect to the degree of Co coverage and the number of Co layers.

Finally, the formation energies of adsorbed oxygen atoms depending on the degree of Co coverage and the number of Co layers are examined in Fig. 7 (bottom). Since the hexagonal lattice of graphene has almost the same lattice parameters as an

*hcp*-Co lattice, the graphene should guide the initial stage of the Co-nanostructure formation. The systems studied here show that planar clusters are initially present only at very thin Co coverage and then change from 2D to 3D Co structures when layers are present. It is clear that with the growth of planar Co clusters below a single layer, the oxidation process is energetically favorable, as well as full oxidation of the cluster (for 0.02-0.06 nm of Co corresponding to < 43% coverage). In contrast to the Co monolayer, Co clusters can easily change their morphology under oxidation and move to 3D systems such as Co oxides as illustrated in Figs. 1b and 1c. We note that the size of our planar Co clusters is limited by the size of the supercell used while, in experimental Co/Gr systems, planar clusters of larger size can be present. During the oxidation process, the morphology of planar clusters may change differently from that is represented here, but the energetics of this oxidation process is similar. The oxidation of a Co monoatom (Fig. 1a) is less energetically favorable than the Co cluster because after the addition of oxygen molecule, it preferentially forms a $CoO_2$ molecule on the graphene surface. With increasing number of Co layers, the joining of the first oxygen molecule becomes more energetically unfavorable, but still remains an exothermic process. The oxidation remains exothermic until full oxidation of the top Co layer is reached at the Co thickness of 1 nm when the process becomes endothermic still with a small formation energy. Therefore, the oxidation process is possible even at room temperature without additional energy. In experimental cobalt layers on graphene there could be defects and grain boundaries that would also decrease the formation energy of oxidation. This suggests that the oxidation process of Co/Gr/Cu will be a self-limited process which should stop after formation of a few layers of CoO over the Co/Gr/Cu composite. As shown in Fig. 6, the degree of oxidation can be relatively high where the ratio of oxidized Co to metallic Co in Co(2 nm)/Gr/Cu is close to 7:3. However, it should be noted that Co oxidation takes place not only on the surface but also at the grain boundaries of Co layers, and the DFT calculations do not take this into account. In this case, a volume of Co would remain unoxidized, which may be explained by a combination of the following effects:

1) The presence of the graphene substrate reduces the gaps between the Co grains because the adsorption of Co on graphene is determined by the morphology of substrate. The Co clusters will prefer to aggregate at impurity sites rather than the pristine graphene surface.

2) Additional calculations for the oxidation of 1, 2, 4 and 6 layers of Co without Gr/Cu substrate demonstrate that oxidation of Co layers without Gr/Cu is energetically more favorable (0.2~0.4 eV/$O_2$) without graphene. Therefore, the presence of Gr/Cu substrate decreases chemical activity of all sites of Co.

On the other hand, the properties of CoO layers in the CoO/Co/Gr/Cu system were further investigated using the LDA+U method [25] for various values of on-site Coulomb repulsion for Co atoms in the CoO-layer ($3d$ orbitals). Results of these calculations (Fig. 8) demonstrate that in contrast to pure oxides of $3d$ metals [25], a layered structure with metal oxide(CoO)/metal(Co)/semi-metal(Gr)/metal(Cu), presents strong hybridization between oxide and metallic layers which dramatically changes the physical properties of oxide. This effect may be interesting for further investigation concerning physical and chemical applications of metal-oxide/metal/graphene composites.

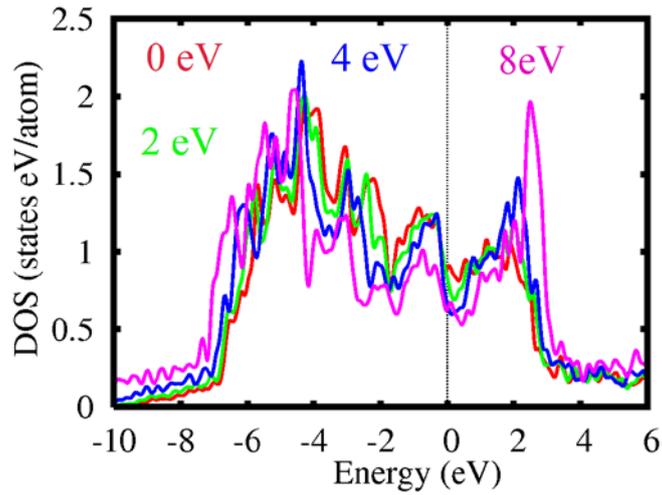

**FIG. 8.** Calculated PDOS of Co atoms from the CoO monolayer in a CoO/Co(5 layers)/Gr/Cu system for various values of on-site Coulomb repulsion. The value of 0 eV corresponds to an LDA calculation.

4. **CONCLUSIONS**

In summary, a series of Co/graphene/Cu composites with tunable Co nanoparticle coverage have been investigated using X-ray spectroscopy and DFT calculation methods. The spectroscopic measurements and DFT calculations suggest that at 0.02 nm and 0.06 nm of Co deposition the Co atoms aggregate on the surface of graphene/Cu as islands. These Co islands are immediately oxidized after exposure to ambient air, forming almost exclusively $Co^{2+}$ species in the form of CoO. For a Co thickness of 2 nm, the upper Co layers (70% of total thickness) are fully oxidized into $Co[OH]_2$ as well as CoO, whereas the lower layers (30%) remain metallic. It is also theoretically predicted that for two or more layers of Co, atoms beneath the oxidized CoO layer display ferromagnetic properties. The preferential formation of CoO at lower Co thickness translates to a layered structure at higher thickness where Co oxide layers protect the inner Co contacting the graphene, allowing it to remain metallic. These results may be useful for spintronics applications requiring either AFM islands or a FM layer. At low Co thickness, AFM CoO islands are preferentially formed that could undergo magnetic exchange interactions to spin polarize the graphene density of states and open a band gap.

**Acknowledgements**

The XPS measurements were supported by the grant of the Russian Scientific Foundation (Project No. 14-22-00004). D.W.B. acknowledges support of DFT calculations from the Ministry of Education and Science of the Russian Federation, Project N 16.1751.2014/K G.S. Chang gratefully acknowledges support from the Natural Sciences and Engineering Research Council of Canada (NSERC) and Canada Foundation for Innovation (CFI).

# REFERENCES

1. J. Wu, D. Zhang, Y. Wang, Y. Wan and B. Hou. *J. Power Sources.* 2012, **198**, 122–126.

2. Y. Liang, H. Wang, J. Zhou, Y. Li, J. Wang, T. Regier, and H. Dai. *J. Am. Chem. Soc.* 2012, **134**, 3517−3523.

3. V. Stamenkovic, B. S. Mun, K. J. J. Mayrhofer, P. N. Ross, N. M. Markovic, J. Rossmeis, J. Greeley and J.K. Norskov. *Angew. Chem., Int. Ed.* 2006, **45**, 2897−2901.

4. M. Hamdani, R. N. Singh and P. Chartier, *Int. J. Electrochem. Sci.* 2010, **5**, 556−577.

5. B. D. Terris, B. D. and T. Thomson. *J. Phys. D.* 2005, **38**, R199–R222.

6. B. Warne, O. I. Kasyutich, E. L. Mayes, J. A. L. Wiggins, and K. K. W. Wong. *IEEE Trans. Magn.* 2000, **36**, 3009-3011.

7. W. S. Leong, H. Gong and J. T. L. Thong. *ACS Nano.* 2014, **8**, 994-1001.

8. K. Nagashio, T. Nishimura, K. Kita and A. Toriumi. *Appl. Phys. Lett.* 2010, **97**, 143514.

9. H. Chen, Q. Niu, Z. Zhang and A. H. MacDonald, *Phys. Rev. B.* 2013, **87**, 144410.

10. R. Decker, J. Brede, N. Atodiresei, V. Caciuc, S. Bluge, and R. Wiesendanger. *Phys. Rev. B.* 2013, **87**, 041403(R).

11. S. R. Power and M. S. Ferreira. *Crystals.* 2013, **3**, 49-78.

12. G. H. Han, F. Gűneş, J. J. Bae, E. S. Kim, S. J. Chae, H. J. Shin, J. Y. Choi, D. Pribat and Y. H. Lee, *Nano Lett.* 2011, **11**, 4144-4148.

13. J. M. Soler, E. Artacho, J. D. Gale, A. Garsia, J. Junquera, P. Orejon and D. Sanchez-Portal, *J. Phys.: Condens. Matter.* 2002, **14**, 2745-2779.

14. D. W. Boukhvalov, Y. N. Gornostyrev, M. A. Uimin, A. V. Korolev and A. Y. Yermakov *RSC Adv.* 2015, **5**, 9173.

15. A. V. Erokhin, E. S. Lokteva, A. Ye. Yermakov, D. W. Boukhvalov, K. I. Maslakov, E. V. Golubina and M. A. Uimin. *Carbon.* 2014, **74**, 291-301.

16. J. P. Perdew and A. Zunger, *Phys. Rev. B*. 1981, **23**, 5048-5079.

17. H. J. Monkhorst and J. D. Pack, *Phys. Rev. B*. 1976, **13**, 5188-5192.

18. D. Yang, A.Velamakanni, G. Bozoklu, S. Park, M. Stoller, R. D. Piner, S. Stankovich, I. Jung, D. A. Field, C. A. Ventrice Jr. and R. S. Ruoff. *Carbon*. 2009, **47**, 145-152.

19. A. Hunt, D. A. Dikin, E. Z. Kurmaev, T. D. Boyko, P Bazylewski, G. S. Chang, A. Moewes. *Adv. Func. Mater.* 2012, **22**, 3950–3957.

20. D. Barreca, A. Gasparotto, O. I. Lebedev, C. Maccato, A. Pozza, E. Tondello, S.Turnerc G.Van Tendelooc. *CrystEngComm*. 2010, **12**, 2185–2197.

21. T. Hofmann, T. H. Yu, M. Folse, L. Weinhardt, M. Bar, Y. Zhang, B. V. Merinov, D. J. Myers, W. A. Goddard and C. Heske, *J. Phys. Chem. C*. 2012, **116**, 24016.

22. C. D. Wagner, L. E. Davis, J. F. Moulder and G. E. Mullenberg. 1978 Handbook of X-ray Photoelectron Spectroscopy (Minnesota: Perkin-Elmer Corporation), 1978.

23. A. Lu, Y. Chen, D. Zeng, M. Li, Q. Xie, X. Zhang and D.-L. Peng. *Nanotechnology.* 2014, **25**, 035707.

24. M. C. Biesinger, B. P. Payne, A. P. Grosvenor, L. W. M. Lau, A. R. Gerson and R. C. Smart. *Appl. Surf. Sci*. 2011, **257**, 2717-2730.

25. V. I. Anisimov, F. Aryasetiawan and A. I. Lichtenstein *J. Phys.: Condens. Matter.* 1997, **9**, 767.